# Direct Measurement of a Non-Hermitian Topological Invariant in a Hybrid Light-Matter System

Rui Su[1,*,†], Eliezer Estrecho[2,*,†], Dąbrówka Biegańska[2,3], Yuqing Huang[1], Matthias Wurdack[2], Maciej Pieczarka[2,3], Andrew G. Truscott[4], Timothy C.H. Liew[1,5], Elena A. Ostrovskaya[2,*] Qihua Xiong[6,7,8,*]

**Affiliations**

[1]Division of Physics and Applied Physics, School of Physical and Mathematical Sciences, Nanyang Technological University, Singapore 637371, Singapore
[2]ARC Centre of Excellence in Future Low-Energy Electronics Technologies and Nonlinear Physics Centre, Research School of Physics, The Australian National University, Canberra 2601, Australia
[3]Department of Experimental Physics, Faculty of Fundamental Problems of Technology, Wrocław University of Science and Technology, Wyb. Wyspiańskiego 27, 50-370 Wrocław, Poland
[4]Laser Physics Centre, Research School of Physics, The Australian National University, Canberra 2601, Australia
[5]MajuLab, International Joint Research Unit UMI 3654, CNRS, Université Côte d'Azur, Sorbonne Université, National University of Singapore, Nanyang Technological University, Singapore
[6]State Key Laboratory of Low-Dimensional Quantum Physics and Department of Physics, Tsinghua University, Beijing 100084, P.R. China
[7]Beijing Academy of Quantum Information Sciences, Beijing 100193, P.R. China
[8]Beijing Innovation Center for Future Chips, Tsinghua University, Beijing, 100084, P.R. China
[†]These authors contributed equally to this work

[*]Corresponding author. Email: Qihua_xiong@tsinghua.edu.cn (Q.X.); elena.ostrovskaya@anu.edu.au (E.A.O.); surui@ntu.edu.sg (R.S.); eliezer.estrecho@anu.edu.au (E.E.)

# ABSTRACT

**Topology is central to understanding and engineering materials that display robust physical phenomena immune to imperfections. Different topological phases of matter are characterised by topological invariants. In energy-conserving (Hermitian) systems, these invariants are determined by the winding of eigenstates in momentum space. In non-Hermitian systems, a novel topological invariant is predicted to emerge from the winding of the complex eigenenergies. Here, we directly measure the non-Hermitian topological invariant arising from exceptional points in the momentum-resolved spectrum of exciton polaritons. These are hybrid light-matter quasiparticles formed by photons strongly coupled to electron-hole pairs (excitons) in a halide perovskite semiconductor at room temperature. We experimentally map out both the real (energy) and imaginary (linewidth) parts of the spectrum near the exceptional points and extract the novel topological invariant - fractional spectral winding. Our work represents an essential step towards realisation of non-Hermitian topological phases in a condensed matter system.**

**Introduction**

Discovery of topologically protected energy bands and associated topological phases in electronic materials have led to demonstrations of unique phenomena, such as dissipationless current (*1*) and enhanced sensitivity to electromagnetic fields (*2, 3*), that have the potential to revolutionise the electronics industry. Inspired by the discoveries in the field of condensed matter physics, the realisation of topological effects in engineered photonic systems holds similar promise for photonic applications (*4*). On the other hand, growing understanding of the physics of non-Hermitian systems with gain and loss (*5, 6*), has led to demonstration of novel functionalities, such as loss-induced lasing (*7*), enhanced sensing (*8, 9*), and optical nonreciprocity (*10, 11*). The last few years have witnessed the convergence of the two research directions, with significant theoretical and experimental advances in extending the notion of topology to non-Hermitian systems (*12, 13*). The bulk-boundary correspondence, the principle relating the surface states to the topological classification of the bulk, was generalised to non-Hermitian systems (*14-17*) and has been explored for high-order systems (*18, 19*). Furthermore, the associated non-Hermitian skin effect, the localisation of bulk modes at the edges of an open boundary system, was observed in experiments (*20-22*). More importantly, a unique non-Hermitian topology arising from the winding of the complex eigenvalues (eigenenergies) was theoretically predicted (*23-25*). This is in stark contrast to energy-conserving systems, where the topological invariants are determined by the winding of the phase of the eigenstates in momentum space, which has been directly measured in ultracold atomic (*26, 27*), microwave (*28*), and photonic systems (*29*). The properties of the eigenstates stemming from the novel non-Hermitian topology, such as the polarisation half charge (*30*) and localisation of modes (*31*), have been experimentally observed in photonic and mechanical systems. However, a direct measurement of the non-Hermitian topological invariant in momentum space is yet to be demonstrated, regardless of the physical nature of the system under investigation.

Exciton polaritons, hybrid light-matter particles arising from strong coupling of confined photons to excitons in a semiconductor, offer a promising platform for investigations of topology and non-Hermitian physics in condensed matter. Artificial lattice potentials (*32-34*) enable exciton polaritons to emulate topological quantum matter (*35*), although the topological gap only opens in very strong magnetic fields requiring a superconducting magnet and cryogenic temperatures. Under similar extreme conditions, exciton-polariton systems also enable a direct measurement of physical quantities directly related to topology, such as the quantum geometric tensor (*36*), including the non-zero Berry curvature (*36-38*). Moreover, due to the photonic and excitonic losses, exciton polaritons are inherently non-Hermitian. A non-Hermitian spectral degeneracy – an exceptional point (EP) (*7, 39*), where both the eigenvalues and eigenvectors coalesce, was demonstrated in exciton-polariton systems (*40, 41*) in parameter space. Since then, new proposals have emerged combining topology and non-Hermiticity of the system using artificial lattices (*42-44*). However, there are no experimental studies yet demonstrating the novel topology arising from non-Hermiticity in exciton-polariton systems.

In this work, we exploit exciton polaritons formed in an optically anisotropic lead halide perovskite crystals embedded in an optical microcavity, to demonstrate the emergence of non-Hermitian topology in an exciton-polariton system at room temperature. First, we develop a non-Hermitian model for the two states of exciton-polaritons pseudospin that accounts for the inherent losses in the system. The exciton-polariton pseudospin originates from the two allowed projections of its spin on the structure axis, and is directly related to the polarisation of the exciton-polariton emission, i.e., cavity photoluminescence (PL) (*45*). The model predicts the formation of two paired EPs in momentum space connected by the topologically protected bulk Fermi arcs (*30*). We also demonstrate theoretically that the topologies of the eigenstates (polarisation winding) and the eigenenergies (spectral winding) are not equivalent, and the former can persist when the latter is absent. Then, by performing spectroscopic measurements of exciton-polariton PL, we experimentally confirm the existence of paired EPs and Fermi arcs linking them in momentum-resolved spectrum. Moreover, the non-Hermiticity results in the appearance of circular polarisation, maximised near the EPs (*46*), which arises from the imaginary part of the artificial in-plane magnetic field acting on the exciton-polariton pseudospin. Most importantly, we provide a direct measurement of the novel non-Hermitian topological invariant – spectral winding – in a spatially homogenous system and demonstrate the topological stability of the EPs and Fermi arcs against the gap-opening perturbations.

**Results**

**Non-Hermitian model for exciton-polariton dispersion.** Losses are unavoidable in exciton-polariton systems due to the finite lifetimes of the cavity photons and excitons. These losses can be fully accounted for using non-Hermitian framework, where both exciton and photon resonances in the cavity are described by complex energies $\tilde{E} = E - i\gamma$, with the real part corresponds to the resonance energy and the imaginary part to the linewidth (inverse lifetime). We therefore model the complex exciton-polariton dispersion (spectrum in momentum space) by using a 4×4 non-Hermitian Hamiltonian describing the coupling of the two polarisation modes of the cavity photons to the corresponding modes of the exciton (*47*):

$$H(\boldsymbol{k}) = \begin{pmatrix} H_c(\boldsymbol{k}) & V\mathbf{1}_{2\times 2} \\ V\mathbf{1}_{2\times 2} & \tilde{E}_x \mathbf{1}_{2\times 2} \end{pmatrix} \quad (1)$$

where $\hbar \boldsymbol{k}$ is the in-plane momentum, $\mathbf{1}_{2\times 2}$ is the 2×2 identity matrix, $\tilde{E}_x = E_x - i\gamma_x$ is the complex exciton energy, and $V$ is the exciton-photon coupling strength. For simplicity, we assume that the exciton spins are degenerate in energy and the coupling strength is isotropic. We model the cavity photon by extending the Hermitian Hamiltonian (*48, 49*) to properly account for the losses:

$$H_c(\boldsymbol{k}) = \begin{pmatrix} \tilde{E}_c(k) & \tilde{\alpha} + \tilde{\beta}(k)e^{-2i\phi} \\ \tilde{\alpha} + \tilde{\beta}(k)e^{2i\phi} & \tilde{E}_c(k) \end{pmatrix} \quad (2)$$

where $\phi$ is the in-plane propagation angle, $\tilde{E}_c(\boldsymbol{k}) = \tilde{E}_{c0} + \tilde{\chi}(k)$ is the mean complex energy of the cavity photon, $\tilde{\chi}(k)$ is a function related to the effective mass (real part) and the momentum-dependent loss rates (imaginary part), $\tilde{\alpha}$ describes the complex energy splitting due to X-Y splitting, and $\tilde{\beta}(k)$ describes TE-TM splitting. The X-Y splitting can arise from the birefringence in the cavity medium (*48, 49*), for example, due to the anisotropic orthorhombic crystal structure of perovskites at room temperature (*33, 50*), which leads to different cavity lengths for the ordinary and extraordinary waves and results in the splitting of both energies and linewidths at normal incidence ($k = 0$). The transverse-electric transverse-magnetic (TE-TM) splitting naturally arises from the polarisation-dependent reflectivity of the dielectric mirrors at oblique angles, inducing an effective spin-orbit coupling (*48-50*) that increases with the angle of incidence (or $k$). The resulting energy splitting is sensitive to the position of the cavity resonance with respect to the distributed Bragg reflector (DBR) stopband but the linewidth consistently increases (decreases) with momentum for TE (TM) modes (*47*). The model Eqs. (1-2) are derived by extracting the resonances in 4×4 transfer matrix simulations (*46*) (see Methods). The behaviour of the energies and linewidths are presented in the Supplementary Materials.

In the strong coupling regime, the model Eqs. (1-2) result in four exciton-polariton branches (see Supplementary Materials). In this work, we focus on the two lower polariton branches since the upper branches are not visible in PL experiments. The lower polaritons at lower momenta $k$ can be described by a model similar to Eq. (2). However, the effective X-Y and TE-TM splitting parameters $\tilde{\alpha}$ and $\tilde{\beta}$ now also depend on the exciton-photon coupling strength $V$ and exciton-photon detuning $\tilde{E}_{c0} - \tilde{E}_x$. In experiments, the exciton-photon coupling strength is typically fixed but the exciton-photon detuning can be varied across the sample due to distinct cavity lengths. The effective 2×2 Hamiltonian can be recast in a more convenient form as $\tilde{E} = \langle \tilde{E}(\boldsymbol{k}) \rangle \mathbf{1}_{2\times 2} + \boldsymbol{G}(\boldsymbol{k}) \cdot \boldsymbol{\sigma}$, where $\langle \tilde{E}(\boldsymbol{k}) \rangle$ is the mean lower polariton complex energy, $\boldsymbol{\sigma} = [\sigma_x, \sigma_y, \sigma_z]^T$ is a vector of Pauli matrices and

$$\boldsymbol{G}(\boldsymbol{k}) = [\tilde{\alpha} + \tilde{\beta}(k)\cos 2\phi, \tilde{\beta}(k)\sin 2\phi, 0] \quad (3)$$

is the effective non-Hermitian gauge field. The complex spectrum can be written as $\tilde{E}_\pm - \langle \tilde{E} \rangle = \pm\sqrt{G_R^2 - G_I^2 + 2i\boldsymbol{G}_R \cdot \boldsymbol{G}_I}$, where $G_R$ and $G_I$ are the real and imaginary parts of the gauge field, respectively. In the Hermitian limit of negligible losses, the spectrum (energy eigenvalues) of the Hamiltonian with gauge field Eq. (3) features two Dirac cones in momentum space, as shown by the energy surfaces in Fig. 1 (A and B). This loss-less approximation has been successfully used to describe several experiments in exciton-polariton systems such as the optical spin-Hall effect (*45*), anomalous Hall effect (*36*), and the measurement of the quantum geometric tensor (*36-38*).

Adding a real-valued $\sigma_z$-component to the gauge field, Eq. (3), *e.g.*, by inducing a Zeeman shift of the exciton energies by applying an out-of-plane magnetic field, would remove the Hermitian degeneracies at the Dirac point and open a topological gap (*36-38*). When polarisation-dependent losses are non-negligible, the imaginary parts of the gauge field, Eq. (3), split each of the Dirac point into a pair of EPs, as shown in Fig. 1 (C and D). These EP pairs are topologically stable (*12*), in stark contrast to the Dirac points that are only stable when protected by symmetry. It takes a strong gap-opening perturbation (i.e. a real-valued $\sigma_z$-term) to make the EPs approach each other (see Fig. 1E), and annihilate to open the gap (see Fig. 1F). A closer look at one of the pairs, as shown in Fig. 1 (C and D), reveals that the paired EPs are connected by open arcs called the bulk Fermi arc (*30*), where $\Delta E = 0$ (green), and the imaginary Fermi arc, where $\Delta \gamma = 0$ (orange), which form closed contours in momentum space (see Fig. 1G). The gap opens when the bulk Fermi arc shrinks and disappears, and the imaginary Fermi arc closes.

Non-Hermitian systems are characterised by two, non-equivalent types of topological winding numbers: the first one is a topological charge of the eigenstates (or pseudospin) arising around singularities in momentum space, and the second one is the non-Hermitian topological charge associated with eigenenergies. For the case of Eq. (3), there are pairs of singularities in the pseudospin texture, around which the in-plane pseudospin component winds. As shown in Fig. 1H for the upper eigenstate, the in-plane pseudospin angle rotates by $\pm \pi$ around the singularity due to the $\pi$-discontinuity at the bulk Fermi arc, resulting in half-integer topological charges (*30*). The lower eigenstate exhibits the same topological charges at the same singularities (see Supplementary Materials). Moreover, the pseudospin is polarised up or down at these points, which translates to circularly polarised PL emission (*51*), exactly at the momenta of the paired EPs (see Supplementary Materials).

Adding a chiral (or $\sigma_z$) term to Eq. (3), which can be achieved by magnetically-induced Zeeman splitting (*36*), or intrinsic chirality (optical activity) (*38*), moves the EPs in momentum space but the pseudospin singularities remain at the same momenta, a phenomenon closely related to the haunting theorem in singular optics (*51*). Interestingly, the singularities reside in separate eigenstates and the topological charge becomes integer-valued. These effects are demonstrated in Fig. 1I for a weak, real-valued $\sigma_z$ perturbation, where one singularity disappears, since it migrates to the other eigenstate (see Supplementary Materials), and the winding of the remaining one is -

$2\pi$. The integer topological charges persist even if the gap opens. Moreover, with increasing magnitude of the $\sigma_z$-term, the polarisation at the EP becomes elliptical, and the discontinuity at the bulk Fermi arc continuously decreases towards zero where the gap opens, as shown in Fig. 1J (see also Supplementary Materials). The half-integer winding, shared by the two eigenstates, is therefore unstable against any $\sigma_z$-perturbation, where a nonzero $\sigma_z$-term suddenly switches the winding from $\pi$ to $2\pi$ (or to zero for the other eigenstate). This transition is reminiscent of the Hermitian case, where the $\sigma_z$-perturbation immediately destroys the Dirac point and opens the gap.

In contrast to the topology of the eigenstates described above, the winding of the eigenenergies is deeply tied to the exceptional point and is topologically stable. The topology is revealed by the "spectral phase" of the difference of the two complex energies $Arg(\Delta\tilde{E})$ (23-25). As shown in Fig. 1 (K and L), the singularities of the spectral phase occur exactly at the EP with a $\pi$-spectral phase winding or half-integer topological charge, regardless of where the singularities of the eigenstates are located in momentum space. This is because the spectral phase jump at the bulk Fermi arc remains equal to $\pi$. These two features, the $\pi$-winding and the $\pi$-phase jump, persist even under a weak, real-valued $\sigma_z$ perturbation, as shown in Fig. 1L. This is in contrast with the behaviour of the eigenstates, where the $\pi$ winding suddenly switches to $2\pi$ (Fig. 1H and 1I) and the phase jump across the bulk Fermi arc decreases with a $\sigma_z$-perturbation (see Supplementary Materials).

At sufficiently strong $\sigma_z$-perturbation, the EPs meet and annihilate, the gap fully opens, and the non-Hermitian topological charges disappear but the topological charges of the eigenstates (polarisation vortices) remain (see Fig. 1J). This demonstrates that the topologies of the eigenstates and the eigenenergies are separable, and measuring the topology of the eigenstates, in general, is not equivalent to measuring the non-Hermitian topology of the eigenenergies. In the following, we experimentally observe paired EPs in an exciton-polariton system with weak chirality and directly measure the non-Hermitian topological invariant by extracting the winding of complex eigenenergies from the PL spectrum.

**Experimental observation of paired EPs.** To demonstrate the EP pairs in the exciton-polariton dispersion experimentally, we employ the microcavity schematically shown in Fig. 2A. It is formed by sandwiching a ~142-nm thick $CsPbBr_3$ perovskite crystal between two $SiO_2/Ta_2O_5$

DBRs, as detailed in the Methods. The crystal is optically anisotropic due to its orthorhombic symmetry (*33, 52, 53*), which leads to X-Y splitting of the exciton-polariton states (*33, 50*). The exciton polaritons are excited by an off-resonant laser with the photon energy far above the perovskite exciton energy. The exciton-polariton energy distribution in momentum space is extracted from the PL of the sample. An emitted photon at polar angle $\theta$, azimuthal angle $\phi$ (see schematics in Fig. 2B), and photon wavelength $\lambda$ carries the exciton-polariton in-plane momentum $\hbar \boldsymbol{k} = \hbar \left(\frac{2\pi}{\lambda}\right) \sin\theta \,(\cos\phi, \sin\phi)$, with $\phi$ corresponding to the propagation angle. To distinguish between the pseudospin states of exciton polaritons, which translate to the polarisation of the PL, the signal is recorded with linear polarisations along the horizontal-vertical (H-V) (orientation shown in Fig. 2B), diagonal-antidiagonal (D-A), and left-right circular polarisations (L-R). The sample is oriented so that the X-Y splitting, along with the spin-orbit coupling, result in energy crossing along $(k_x, k_y = 0)$ but no crossing along $(k_x = 0, k_y)$ in the linearly polarised exciton-polariton dispersions (*36, 37*), as shown in Fig. 2C.

The non-Hermitian character of the exciton-polariton dispersion is reflected in the linewidths of the modes, which are also split at $k = 0$ (see fig. S2). Subtracting the mean value, *i.e.*, $\gamma - \langle\gamma\rangle$, reveals that the linewidth dependence on $k$ is also anisotropic as shown in Fig. 2D, such that the linewidth switches or crosses along the direction $(k_x, k_y = 0)$, but not along $(k_x = 0, k_y)$. The crossings in energy and linewidth along the same direction suggest that the Fermi arcs form two loops in momentum space, as shown by the insets of Fig. 2C. A similar behaviour of energy and linewidth in momentum space was observed for the cavity photons in birefringent ZnO-based microcavities (*46, 54*) in the weak coupling regime (i.e. without coupling to excitons). However, the paired EPs remained elusive in the strong coupling regime despite several experiments on exciton polaritons in anisotropic cavities (*36-38, 55*). Related EPs in momentum space were observed in microcavities with embedded carbon nanotubes (*56*) and organic microcrystals (*57*), but strong exciton-photon coupling in these systems only occurs in one polarisation. Our results demonstrate that the exciton polaritons can inherit the EPs from birefringent cavity photons.

The EPs predicted in Figs. 1C and 1D are expected to exist near the energy crossings at $\boldsymbol{k}^* \approx (\pm 5.2, 0)$ μm$^{-1}$ (see Fig. 2C). The position of the EP pair can be determined by carefully tracking the complex spectrum near this region. The extraction of peak energy and linewidth from the polarised PL measurements is detailed in the Methods. Figure 3 shows the results of the measurements along five lines (labelled b-f) in $k$-space that intersect the Fermi arcs as schematically shown in Fig. 3A. The measurement in Fig. 3B is approximately along the bulk Fermi arc, where the mode energies approach each other while the linewidths clearly repel. At a slightly off-arc position, as shown in Fig. 3C, the mode energies always repel, but the linewidths cross at two points of the imaginary Fermi arc. Perpendicular to the bulk Fermi arc and close to the EP, the energies cross while the linewidths approach each other, as shown in Fig. 3D. Conversely, the modes cross in linewidth and approach in energy outside the bulk Fermi arc but close to the EP, as shown in Fig. 3F. Across the middle of the bulk Fermi arc, Fig. 3E clearly shows that the energies cross, but the linewidths repel. From these results (see fig. S5 for the 2D surfaces), we estimate the EP positions to be $\boldsymbol{k_{EP}} \approx (-5.2, 0.40)$ µm$^{-1}$ and $\boldsymbol{k_{EP}} \approx (-5.2, 0.09)$ µm$^{-1}$ with a bulk Fermi arc length of $\approx 0.31$ µm$^{-1}$.

**Pseudospin texture in the complex artificial gauge field.** The existence of the EPs is further evidenced by the circular polarisation of the exciton-polariton emission (*9, 46*), which corresponds to the singularities of the eigenstates near the EPs (see Fig. 1H, I) where the exciton-polariton pseudospin points either up or down. We define the pseudospin of the eigenstates using the Stokes parameters: $S_1 = (I_H - I_V)/(I_H + I_V)$, $S_2 = (I_D - I_A)/(I_D + I_A)$, and $S_3 = (I_R - I_L)/(I_R + I_L)$. In the Hermitian limit, and since Eq. (3) does not have a $\sigma_z$-term, the eigenstates are orthogonal and purely linearly polarised (*48*), with the corresponding pseudospins confined to the $S_1$-$S_2$ plane of the Poincaré sphere (orthogonal polarisations are antipodal), as shown by the thin red and blue arrows in Fig. 4A. However, due to non-Hermiticity, the eigenstates of the Hamiltonian are not orthogonal, and the pseudospins of the eigenstates tend to point in the same direction towards one of the poles, as shown by the thick red and blue arrows in Fig. 4A. This leads to a non-zero $S_3$ Stokes component, while the projections on the $S_1$-$S_2$ plane remain antipodal. Hence, both eigenstates have the same $S_3$ components (dashed arrows in Fig. 4A) which, in this case, is a measure of the non-Hermiticity of the Hamiltonian. At the EP, full alignment occurs, resulting in a merged eigenstate pointing to the pole with a purely circular polarisation, as shown by the purple arrow in Fig. 4A. The calculated circular polarisation or $S_3$ component of the pseudospin texture of either eigenstate in $k$-space is shown in Fig. 4B. Maximum circular polarisation occurs at the EPs and gradually decreases away from them. The EPs within the pair have opposite chirality and the two pairs have opposite orientations.

The appearance of chirality in the model arises from the interplay between the real and imaginary components of the purely in-plane complex artificial magnetic field. If the real and imaginary fields are parallel, or purely real or imaginary, the pseudospin of the eigenstates tend to align (parallel or antiparallel) to the field. However, if the two fields have perpendicular components, the pseudospins tend to align away from the real and imaginary parts and towards each other, which in our case effectively induces an out-of-plane component. The effective out-of-plane component is different from a real-valued out-of-plane magnetic field, where the pseudospins of the two modes remain antipodal on the Poincaré sphere. This non-Hermitian generalisation allows an arbitrary control of the polarisation (*58*) and can lead to rich spin dynamics not achievable with real-valued artificial magnetic fields. Note that in this off-resonant (incoherent) regime of exciton-polariton excitation, we are measuring the pseudospin of the eigenstates. This is in contrast to the resonant (coherent) regime, where a non-zero $S_3$-component can result from pseudospin precession in an in-plane field (*45*).

We take advantage of the non-Hermiticity, which results in non-orthogonal and chiral eigenstates, to directly measure the $S_3$ or spin texture of the exciton polaritons, as shown in Fig. 4C, by capturing the momentum space distribution without resolving the two modes. This method assumes that the two eigenstates at a momentum ***k*** are equally occupied, which is not always the case. However, it is effective for finding the pseudospin singularities shown in Fig. 1(H and I). Indeed, a circular polarisation texture qualitatively similar to the prediction of the model is observed using this method, with the local extrema near the EPs (black points in Fig. 4C). The discrepancy between the momenta of the EPs and the extrema of the $S_3$ texture is due to the close proximity of the EPs. The opposite circular polarisation in the vicinity of the paired EPs tend to overlap and cancel each other. Hence, the measured $|S_3|$ is greatly reduced and the extrema are offset away from the EPs (see Supplementary Materials for supporting simulations). Similar low level of circular polarisation degree near the EPs was observed for microcavity photons without coupling to excitons (*54*).

In addition to the spin texture due to the EPs, there is a background circular polarisation (or chirality) that is not accounted for in the model Eqs. (1,2). This originates from the exciton emission of the bare perovskite (see Supplementary Materials). The observed chirality can arise from the chirality of the excitons in lead-halide perovskites (*59, 60*) but further experimental work is needed to verify its origin and derive an effective model for its spin texture. Here, we treat the chirality as a weak $\sigma_z$ perturbation to Eq. (3) which can move the EPs towards each other and potentially open the gap in the Hermitian limit (*38*) (see Fig. 1). The clear observation of EPs in our experiment therefore demonstrates the topological stability of EP pairs against gap-opening (chiral) perturbations, or any perturbation in general (*12*). More importantly, the weak chirality places the experiment in the regime shown in Fig. 1(E, I, and L), where the topologies of the eigenstates and eigenenergies are not related to each other.

**Observation of non-Hermitian topological invariant.** Finally, with the existence of the EPs verified using both the complex energies and pseudospin texture, we demonstrate the direct measurement of the non-Hermitian topological invariant arising from the EPs in momentum space. For the two-level system considered here, the non-Hermitian topological invariant, called the 'spectral winding' or 'vorticity' (*23-25*) is formally defined as:

$$w = -\frac{1}{2\pi} \oint_C \nabla_k \arg[\tilde{E}_+(\boldsymbol{k}) - \tilde{E}_-(\boldsymbol{k})] \cdot d\boldsymbol{k} \qquad (4)$$

where $C$ is a closed loop in $k$-space. Naturally, this topological invariant is zero for Hermitian systems. The topology depends on the scalar field $\arg[\tilde{E}_+(\boldsymbol{k}) - \tilde{E}_-(\boldsymbol{k})]$, a 'spectral phase' which is well defined everywhere except at the EPs. Hence, the EPs are sources of non-Hermitian topological charges. For the paired EPs considered here, the theoretical spectral phase calculated from model Eq. (3) rotates in opposite directions around each EP, forming oppositely charged spectral vortices, as shown in Fig. 4D. More importantly, the spectral vortices have half-integer charge (*23*) since the spectral phase acquired around the loop enclosing a single EP is $\pm\pi$.

By carefully measuring the energies and linewidths in the vicinity of the EP pairs, we are able to extract the spectral phase, and consequently determine the winding of the complex eigenenergies, as presented in Fig. 4 (E and F). Clearly, the spectral phase winds around the EPs and jumps by approximately $\pi$ at the bulk Fermi arc that connects the EPs. The small phase jumps away from the EP pair are experimental artefacts where we switch between H-V and D-A polarised measurements (see Methods) to extract the energies and linewidths. Using the definition in Eq. (4), we can assign a $\pm 1/2$ non-Hermitian topological charge to the EPs, as annotated in Fig. 4D, and symbolised by the black and white arrows Fig. 4 (E and F). Each pair of EPs therefore forms a "topological dipole", and the two dipoles have opposite orientations, as predicted by the model in Fig. 4B. Furthermore, the spectral winding around the whole EP pair is zero. Consequently, if the separation of the EP pair is not resolved in the experiment, the non-Hermitian topological invariant would not be measurable.

It is important to stress that the topological winding of the eigenenergies measured here should, in principle, be accompanied by the half-integer winding of polarisation (*30*), as theoretically demonstrated in Fig. 1 (H and K). However, due to the background $S_3$, which introduces a weak $\sigma_z$-term perturbation, the measured spectral winding is no longer related to the winding of the polarisation (see Fig. 1 I and L). Hence, the non-Hermitian topological invariant observed in this work is fundamentally different from the winding of the eigenstates observed in photonic systems (*30*). Moreover, the measured half-integer topological invariant is unaffected by the chirality observed in the experiment, as theoretically demonstrated in Fig. 1L. This is not the case for the polarisation winding which would become integer-valued even for weak chiral perturbation.

**Discussion**

In summary, we have demonstrated the existence of paired EPs in the momentum-resolved exciton-polariton spectrum and have directly measured the non-Hermitian topological invariant arising from the half-integer winding of the exciton-polariton complex eigenenergies around the EPs. We have also shown theoretically that the topology of the eigenstates and eigenenergies are separable and hence the signatures of inherent topology of exceptional points previously observed in the eigenstates of classical wave systems (*30, 31*) are fundamentally different from our observation.

In contrast to previously demonstrated EPs in parameter space of exciton-polariton systems(*40, 41, 56*), the EPs in momentum space observed here are expected to have a direct influence on the system's dynamics (*61*). Our observation can potentially lead to the realisation of non-Hermitian topological phases (*24*) and the predicted non-trivial dynamics, such as a non-Hermitian skin effect(*62*), without the need for sophisticated microstructuring of the sample (*43*) or strong external magnetic fields (*35*). Moreover, we have demonstrated the manifestation of the imaginary part of the artificial gauge field that tends to align the exciton-polariton pseudospin pair towards each other and perpendicular to the field direction. This may lead to a new type of spin precession (*58*) and dynamics of exciton polaritons that is not possible in real magnetic fields. Combined with advanced methods for potential landscaping (*32*) and the possibility to extract a wide range of observables from the cavity PL, our work affirms exciton polaritons as a solid-state platform for exploring robust topological phenomena that do not occur in Hermitian systems. Indeed, a recent experiment on organic microcavities with highly polarisation-dependent (anisotropic) light-matter coupling (*57*) has demonstrated a diverging quantum metric at the EP, in stark difference to Hermitian systems (*31*).

Unlike previous observations of exceptional points in optical microcavities (*54*), our demonstration of a non-Hermitian topological invariant relies on hybrid light-matter particles, exciton polaritons, which exhibit strong interactions due to the exciton component (*63*). Therefore, our study offers a new platform for investigating the interplay between the non-Hermitian topology and nonlinearity, which may bring about unexpected phenomena, such as, e.g., similar to self-adaptation in energy transfer (*64*). For example, under a strong circular polarized excitation, the unique strong spin-anisotropic nonlinearity in exciton- polariton systems (*48, 55*) could potentially lead to an effective real $\sigma_z$-perturbation with rich tunability. This can could provide an efficient pathway for investigating Hermitian and non-Hermitian topological effects in the presence of $\sigma_z$-perturbations and nonlinearity, even without the need of real magnetic fields.

Finally, the strong interactions and very small effective mass of exciton polaritons has successfully enabled demonstration of collective quantum effects, e.g., bosonic condensation (*65*) and superfluidity (*66, 67*), at elevated temperatures, in particular, using lead-halide perovskites (*33, 34, 68*). Thus, our work paves the way for investigating the interplay between quantum many-body effects and non-Hermitian topology, which is as yet an unexplored frontier in non-Hermitian physics (*12*).

## MATERIALS AND METHODS

**Perovskite microcavity fabrication**

20.5 pairs of $SiO_2$ and $Ta_2O_5$ were deposited on a silicon substrate as the bottom DBR using an electron beam evaporator (OHMIKER-50D). The 142 nm-thick cesium lead bromide perovskite crystal was grown with a vapor phase deposition method on a mica substrate and then transferred onto the bottom DBR by a dry-transfer process with scotch tape(*33*). Subsequently, a 60-nm thick Poly(methyl methacrylate) protection layer was spin-coated onto the perovskite layer. Another 10.5 pairs of $SiO_2$ and $Ta_2O_5$ were deposited onto the structure by the e-beam evaporator, acting as the top DBR to complete the fabrication process.

**Optical spectroscopy characterisations**

The energy-resolved momentum-space PL was mapped by using a home-built angle-resolved PL setup with a motorised translation stage in order to scan the whole 2D-momentum space. In the detection line, a quarter-wave plate, a half-wave plate and a linear polariser were used for the detection of polarisation-resolved PL mappings in momentum space. A continuous-wave laser (457 nm) with a pump spot of ~10 μm was used to pump the perovskite microcavity, passing through an optical chopper to minimise sample heating. The emission from the perovskite microcavity was collected through a 50× objective (NA= 0.75, Mitutoyo), and directed to a 550-mm focal length spectrometer (HORIBA iHR550) with a grating of 1200 lines/mm and a liquid nitrogen–cooled charge coupled device of 256×1024 pixels. All measurements were conducted at room temperature.

**Non-Hermitian theoretical model**

The simple non-Hermitian model in Eq. (1,2) for the exciton-polariton spectrum was derived by simulating the reflectance of a microcavity with an embedded anisotropic cavity spacer and the excitonic transition in the strong coupling regime. We follow the 4×4 transfer matrix method of Ref.(*46*) but with an addition of the exciton resonance modelled as a Lorentz oscillator.

The transfer matrix calculations and the theoretical model also capture the linewidth behaviour of the experiment shown in fig. S4. Regardless of the direction, the linewidth increases with $k$ as the exciton fraction of polariton increases. However, the experimental linewidth increases more or less linearly with $k$ (see fig. S4), compared to the near parabolic behaviour of the numerical simulation.

This can arise from the inhomogeneous broadening of the exciton resonance, which is not accounted for in the simulations.

**Determination of mode energies and linewidths**

To measure the energy and linewidth, we fit Lorentzian functions to the measured spectra at different points in $k$-space. The energy corresponds to the centre while the linewidth corresponds to the full-width-at-half-maximum of the fitted Lorentzian function. Away from the energy crossings, the spectrum displays two peaks and can be fitted with a double Lorentzian function, as shown in fig. S7. Near the energy crossings, there is only one apparent peak since the mode energy separation is smaller than the linewidth. To resolve the individual peaks, we take advantage of the orthogonal pairs (H-V or D-A) of polarised measurements. Each polarised spectrum is fitted with a single Lorentzian as shown in fig. S7 and the orthogonal pair with the largest energy splitting is chosen. This switching between H-V and D-A results in jumps in the extracted energies and linewidths (see Fig. 3B-F) and small phase jumps in the spectral phase (see Fig. 4E-F). This is because we are not measuring (or projecting) the eigenstates in their appropriate orthogonal basis. In principle, a full polarisation tomography is needed, in addition to the 2D scan of the momentum space, to properly separate the modes and smoothen the jumps in the complex energy and spectral phase. However, this will greatly increase the measurement time and data from 3D to 4D. The current set of data is enough to verify the existence of EPs and measure the half-integer spectral winding in this system.

# REFERENCES AND NOTES:

1. Y. Tokura, K. Yasuda, A. Tsukazaki, Magnetic topological insulators. *Nat Rev Phys* **1**, 126-143 (2019).
2. A. A. Burkov, Chiral anomaly and transport in Weyl metals. *J Phys-Condens Mat* **27**, (2015).
3. C. K. Chan, N. H. Lindner, G. Refael, P. A. Lee, Photocurrents in Weyl semimetals. *Phys Rev B* **95**, (2017).
4. L. Lu, J. D. Joannopoulos, M. Soljacic, Topological photonics. *Nat Photonics* **8**, 821-829 (2014).
5. Ş. K. Özdemir, S. Rotter, F. Nori, L. Yang, Parity–time symmetry and exceptional points in photonics. *Nature Materials* **18**, 783-798 (2019).
6. R. El-Ganainy, K. G. Makris, M. Khajavikhan, Z. H. Musslimani, S. Rotter, D. N. Christodoulides, Non-Hermitian physics and PT symmetry. *Nat. Phys.* **14**, 11-19 (2018).
7. B. Peng, Ş. K. Özdemir, S. Rotter, H. Yilmaz, M. Liertzer, F. Monifi, C. M. Bender, F. Nori, L. Yang, Loss-induced suppression and revival of lasing. *Science* **346**, 328-332 (2014).
8. W. Chen, Ş. Kaya Özdemir, G. Zhao, J. Wiersig, L. Yang, Exceptional points enhance sensing in an optical microcavity. *Nature* **548**, 192-196 (2017).
9. H. Hodaei, A. U. Hassan, S. Wittek, H. Garcia-Gracia, R. El-Ganainy, D. N. Christodoulides, M. Khajavikhan, Enhanced sensitivity at higher-order exceptional points. *Nature* **548**, 187-191 (2017).
10. C. E. Rüter, K. G. Makris, R. El-Ganainy, D. N. Christodoulides, M. Segev, D. Kip, Observation of parity–time symmetry in optics. *Nat. Phys.* **6**, 192-195 (2010).
11. B. Peng, Ş. K. Özdemir, F. Lei, F. Monifi, M. Gianfreda, G. L. Long, S. Fan, F. Nori, C. M. Bender, L. Yang, Parity–time-symmetric whispering-gallery microcavities. *Nat. Phys.* **10**, 394-398 (2014).
12. E. J. Bergholtz, J. C. Budich, F. K. Kunst, Exceptional topology of non-Hermitian systems. *arXiv preprint arXiv:1912.10048*, (2019).
13. A. Ghatak, T. Das, New topological invariants in non-Hermitian systems. *J. Phys. Condens. Matter.* **31**, 263001 (2019).
14. S. Yao, Z. Wang, Edge States and Topological Invariants of Non-Hermitian Systems. *Phys. Rev. Lett.* **121**, 086803 (2018).
15. Y. Xiong, Why does bulk boundary correspondence fail in some non-hermitian topological models. *J. Phys. Commun.* **2**, 035043 (2018).
16. F. K. Kunst, E. Edvardsson, J. C. Budich, E. J. Bergholtz, Biorthogonal Bulk-Boundary Correspondence in Non-Hermitian Systems. *Phys. Rev. Lett.* **121**, 026808 (2018).
17. S. Yao, F. Song, Z. Wang, Non-Hermitian Chern Bands. *Phys. Rev. Lett.* **121**, 136802 (2018).
18. T. Liu, Y.-R. Zhang, Q. Ai, Z. Gong, K. Kawabata, M. Ueda, F. Nori, Second-Order Topological Phases in Non-Hermitian Systems. *Phys. Rev. Lett.* **122**, 076801 (2019).
19. C. H. Lee, L. Li, J. Gong, Hybrid Higher-Order Skin-Topological Modes in Nonreciprocal Systems. *Phys. Rev. Lett.* **123**, 016805 (2019).
20. M. Brandenbourger, X. Locsin, E. Lerner, C. Coulais, Non-reciprocal robotic metamaterials. *Nat. Commun.* **10**, 4608 (2019).
21. T. Helbig, T. Hofmann, S. Imhof, M. Abdelghany, T. Kiessling, L. W. Molenkamp, C. H. Lee, A. Szameit, M. Greiter, R. Thomale, Generalized bulk–boundary correspondence in non-Hermitian topolectrical circuits. *Nat. Phys.* **16**, 747-750 (2020).
22. L. Xiao, T. Deng, K. Wang, G. Zhu, Z. Wang, W. Yi, P. Xue, Non-Hermitian bulk–boundary correspondence in quantum dynamics. *Nat. Phys.* **16**, 761-766 (2020).
23. D. Leykam, K. Y. Bliokh, C. Huang, Y. D. Chong, F. Nori, Edge Modes, Degeneracies, and Topological Numbers in Non-Hermitian Systems. *Phys. Rev. Lett.* **118**, 040401 (2017).

24. Z. Gong, Y. Ashida, K. Kawabata, K. Takasan, S. Higashikawa, M. Ueda, Topological Phases of Non-Hermitian Systems. *Phys. Rev. X* **8**, 031079 (2018).
25. H. Shen, B. Zhen, L. Fu, Topological Band Theory for Non-Hermitian Hamiltonians. *Phys. Rev. Lett.* **120**, 146402 (2018).
26. M. Atala, M. Aidelsburger, J. T. Barreiro, D. Abanin, T. Kitagawa, E. Demler, I. Bloch, Direct measurement of the Zak phase in topological Bloch bands. *Nat. Phys.* **9**, 795-800 (2013).
27. M. Aidelsburger, M. Lohse, C. Schweizer, M. Atala, J. T. Barreiro, S. Nascimbène, N. R. Cooper, I. Bloch, N. Goldman, Measuring the Chern number of Hofstadter bands with ultracold bosonic atoms. *Nat. Phys.* **11**, 162-166 (2015).
28. W. Hu, J. C. Pillay, K. Wu, M. Pasek, P. P. Shum, Y. D. Chong, Measurement of a Topological Edge Invariant in a Microwave Network. *Phys. Rev. X* **5**, 011012 (2015).
29. S. Mittal, S. Ganeshan, J. Fan, A. Vaezi, M. Hafezi, Measurement of topological invariants in a 2D photonic system. *Nat. Photon.* **10**, 180-183 (2016).
30. H. Zhou, C. Peng, Y. Yoon, C. W. Hsu, K. A. Nelson, L. Fu, J. D. Joannopoulos, M. Soljačić, B. Zhen, Observation of bulk Fermi arc and polarization half charge from paired exceptional points. *Science* **359**, 1009-1012 (2018).
31. A. Ghatak, M. Brandenbourger, J. van Wezel, C. Coulais, Observation of non-Hermitian topology and its bulk-edge correspondence in an active mechanical metamaterial. *Proc Natl Acad Sci U S A*, 202010580 (2020).
32. C. Schneider, K. Winkler, M. D. Fraser, M. Kamp, Y. Yamamoto, E. A. Ostrovskaya, S. Höfling, Exciton-polariton trapping and potential landscape engineering. *Rep. Prog. Phys.* **80**, 016503 (2016).
33. R. Su, S. Ghosh, J. Wang, S. Liu, C. Diederichs, T. C. H. Liew, Q. Xiong, Observation of exciton polariton condensation in a perovskite lattice at room temperature. *Nat. Phys.* **16**, 301-306 (2020).
34. M. Dusel, S. Betzold, O. A. Egorov, S. Klembt, J. Ohmer, U. Fischer, S. Höfling, C. Schneider, Room temperature organic exciton–polariton condensate in a lattice. *Nat. Commun.* **11**, 2863 (2020).
35. S. Klembt, T. H. Harder, O. A. Egorov, K. Winkler, R. Ge, M. A. Bandres, M. Emmerling, L. Worschech, T. C. H. Liew, M. Segev, C. Schneider, S. Höfling, Exciton-polariton topological insulator. *Nature* **562**, 552-556 (2018).
36. A. Gianfrate, O. Bleu, L. Dominici, V. Ardizzone, M. De Giorgi, D. Ballarini, G. Lerario, K. W. West, L. N. Pfeiffer, D. D. Solnyshkov, D. Sanvitto, G. Malpuech, Measurement of the quantum geometric tensor and of the anomalous Hall drift. *Nature* **578**, 381-385 (2020).
37. L. Polimeno, M. De Giorgi, G. Lerario, L. De Marco, L. Dominici, V. Ardizzone, M. Pugliese, C. T. Prontera, V. Maiorano, A. Moliterni, Tuning the Berry curvature in 2D Perovskite. *arXiv preprint arXiv:2007.14945*, (2020).
38. J. Ren, Q. Liao, F. Li, Y. Li, O. Bleu, G. Malpuech, J. Yao, H. Fu, D. Solnyshkov, Nontrivial band geometry in an optically active system. *Nat. Commun.* **12**, 689 (2021).
39. W. D. Heiss, Exceptional points of non-Hermitian operators. *Journal of Physics A: Mathematical and General* **37**, 2455-2464 (2004).
40. T. Gao, E. Estrecho, K. Y. Bliokh, T. C. H. Liew, M. D. Fraser, S. Brodbeck, M. Kamp, C. Schneider, S. Höfling, Y. Yamamoto, F. Nori, Y. S. Kivshar, A. G. Truscott, R. G. Dall, E. A. Ostrovskaya, Observation of non-Hermitian degeneracies in a chaotic exciton-polariton billiard. *Nature* **526**, 554-558 (2015).
41. T. Gao, G. Li, E. Estrecho, T. C. H. Liew, D. Comber-Todd, A. Nalitov, M. Steger, K. West, L. Pfeiffer, D. W. Snoke, A. V. Kavokin, A. G. Truscott, E. A. Ostrovskaya, Chiral Modes at Exceptional Points in Exciton-Polariton Quantum Fluids. *Phys. Rev. Lett.* **120**, 065301 (2018).

42. P. Comaron, V. Shahnazaryan, W. Brzezicki, T. Hyart, M. Matuszewski, Non-Hermitian topological end-mode lasing in polariton systems. *Phys. Rev. Research* **2**, 022051 (2020).
43. S. Mandal, R. Banerjee, E. A. Ostrovskaya, T. C. H. Liew, Nonreciprocal Transport of Exciton Polaritons in a Non-Hermitian Chain. *Phys. Rev. Lett.* **125**, 123902 (2020).
44. L. Pickup, H. Sigurdsson, J. Ruostekoski, P. G. Lagoudakis, Synthetic band-structure engineering in polariton crystals with non-Hermitian topological phases. *Nat. Commun.* **11**, 4431 (2020).
45. C. Leyder, M. Romanelli, J. P. Karr, E. Giacobino, T. C. H. Liew, M. M. Glazov, A. V. Kavokin, G. Malpuech, A. Bramati, Observation of the optical spin Hall effect. *Nat. Phys.* **3**, 628-631 (2007).
46. S. Richter, T. Michalsky, C. Sturm, B. Rosenow, M. Grundmann, R. Schmidt-Grund, Exceptional points in anisotropic planar microcavities. *Phys. Rev. A* **95**, 023836 (2017).
47. G. Panzarini, L. C. Andreani, A. Armitage, D. Baxter, M. S. Skolnick, V. N. Astratov, J. S. Roberts, A. V. Kavokin, M. R. Vladimirova, M. A. Kaliteevski, Cavity-polariton dispersion and polarization splitting in single and coupled semiconductor microcavities. *Physics of the Solid State* **41**, 1223-1238 (1999).
48. H. Terças, H. Flayac, D. D. Solnyshkov, G. Malpuech, Non-Abelian Gauge Fields in Photonic Cavities and Photonic Superfluids. *Phys. Rev. Lett.* **112**, 066402 (2014).
49. A. Fieramosca, L. Polimeno, G. Lerario, L. De Marco, M. De Giorgi, D. Ballarini, L. Dominici, V. Ardizzone, M. Pugliese, V. Maiorano, Chromodynamics of photons in an artificial non-Abelian magnetic Yang-Mills field. *arXiv preprint arXiv:1912.09684*, (2019).
50. R. Su, S. Ghosh, T. C. H. Liew, Q. Xiong, Optical switching of topological phase in a perovskite polariton lattice. *Science Advances* **7**, eabf8049 (2021).
51. M. V. Berry, M. R. Dennis, The optical singularities of birefringent dichroic chiral crystals. *Proceedings of the Royal Society of London. Series A: Mathematical, Physical and Engineering Sciences* **459**, 1261-1292 (2003).
52. M. A. Becker, R. Vaxenburg, G. Nedelcu, P. C. Sercel, A. Shabaev, M. J. Mehl, J. G. Michopoulos, S. G. Lambrakos, N. Bernstein, J. L. Lyons, T. Stöferle, R. F. Mahrt, M. V. Kovalenko, D. J. Norris, G. Rainò, A. L. Efros, Bright triplet excitons in caesium lead halide perovskites. *Nature* **553**, 189-193 (2018).
53. R. Su, C. Diederichs, J. Wang, T. C. H. Liew, J. Zhao, S. Liu, W. Xu, Z. Chen, Q. Xiong, Room-Temperature Polariton Lasing in All-Inorganic Perovskite Nanoplatelets. *Nano Lett.* **17**, 3982-3988 (2017).
54. S. Richter, H.-G. Zirnstein, J. Zúñiga-Pérez, E. Krüger, C. Deparis, L. Trefflich, C. Sturm, B. Rosenow, M. Grundmann, R. Schmidt-Grund, Voigt Exceptional Points in an Anisotropic ZnO-Based Planar Microcavity: Square-Root Topology, Polarization Vortices, and Circularity. *Phys. Rev. Lett.* **123**, 227401 (2019).
55. D. Biegańska, M. Pieczarka, E. Estrecho, M. Steger, D. Snoke, K. West, L. Pfeiffer, M. Syperek, A. Truscott, E. Ostrovskaya, Collective excitations of exciton-polariton condensates in a synthetic gauge field. *arXiv preprint arXiv:2011.13290*, (2020).
56. W. Gao, X. Li, M. Bamba, J. Kono, Continuous transition between weak and ultrastrong coupling through exceptional points in carbon nanotube microcavity exciton–polaritons. *Nat. Photon.* **12**, 362-367 (2018).
57. Q. Liao, C. Leblanc, J. Ren, F. Li, Y. Li, D. Solnyshkov, G. Malpuech, J. Yao, H. Fu, Experimental measurement of the divergent quantum metric of an exceptional point. *arXiv preprint arXiv:2011.12037*, (2020).
58. A. Cerjan, S. Fan, Achieving Arbitrary Control over Pairs of Polarization States Using Complex Birefringent Metamaterials. *Phys. Rev. Lett.* **118**, 253902 (2017).
59. P. C. Sercel, Z. V. Vardeny, A. L. Efros, Circular dichroism in non-chiral metal halide perovskites. *Nanoscale* **12**, 18067-18078 (2020).

60. J. Li, J. Li, R. Liu, Y. Tu, Y. Li, J. Cheng, T. He, X. Zhu, Autonomous discovery of optically active chiral inorganic perovskite nanocrystals through an intelligent cloud lab. *Nat. Commun.* **11**, 2046 (2020).
61. D. D. Solnyshkov, C. Leblanc, L. Bessonart, A. Nalitov, J. Ren, Q. Liao, F. Li, G. Malpuech, Quantum metric and wave packets at exceptional points in non-Hermitian systems. *Physical Review B* **103**, 125302 (2021).
62. T. Hofmann, T. Helbig, F. Schindler, N. Salgo, M. Brzezińska, M. Greiter, T. Kiessling, D. Wolf, A. Vollhardt, A. Kabaši, C. H. Lee, A. Bilušić, R. Thomale, T. Neupert, Reciprocal skin effect and its realization in a topolectrical circuit. *Phys. Rev. Research* **2**, 023265 (2020).
63. J. Wu, S. Ghosh, R. Su, A. Fieramosca, T. C. H. Liew, Q. Xiong, Nonlinear Parametric Scattering of Exciton Polaritons in Perovskite Microcavities. *Nano Lett.* **21**, 3120-3126 (2021).
64. S. Assawaworrarit, X. Yu, S. Fan, Robust wireless power transfer using a nonlinear parity–time-symmetric circuit. *Nature* **546**, 387-390 (2017).
65. J. Kasprzak, M. Richard, S. Kundermann, A. Baas, P. Jeambrun, J. M. J. Keeling, F. M. Marchetti, M. H. Szymańska, R. André, J. L. Staehli, V. Savona, P. B. Littlewood, B. Deveaud, L. S. Dang, Bose–Einstein condensation of exciton polaritons. *Nature* **443**, 409-414 (2006).
66. A. Amo, J. Lefrère, S. Pigeon, C. Adrados, C. Ciuti, I. Carusotto, R. Houdré, E. Giacobino, A. Bramati, Superfluidity of polaritons in semiconductor microcavities. *Nat. Phys.* **5**, 805-810 (2009).
67. G. Lerario, A. Fieramosca, F. Barachati, D. Ballarini, K. S. Daskalakis, L. Dominici, M. De Giorgi, S. A. Maier, G. Gigli, S. Kéna-Cohen, D. Sanvitto, Room-temperature superfluidity in a polariton condensate. *Nat. Phys.* **13**, 837-841 (2017).
68. R. Su, J. Wang, J. Zhao, J. Xing, W. Zhao, C. Diederichs, T. C. H. Liew, Q. Xiong, Room temperature long-range coherent exciton polariton condensate flow in lead halide perovskites. *Science Advances* **4**, eaau0244 (2018).

**Acknowledgements:**

**Funding:** Q.X. gratefully acknowledges National Natural Science Foundation of China (No. 12020101003), strong support from the State Key Laboratory of Low-Dimensional Quantum Physics and start-up grant from Tsinghua University. T.C.H.L. acknowledges the support from Singapore Ministry of Education via AcRF Tier 3 Programme "Geometrical Quantum Materials" (MOE2018-T3-1-002), AcRF Tier 2 grants (MOE2017-T2-1-001, MOE2018-T2-02-068 and MOE2019-T2-1-004). E.E., E.A.O., M.W., and M.P. acknowledge support from the Australian Research Council (ARC) through the Centre of Excellence Grant CE170100039. MP also acknowledges support from the Foundation for Polish Science in the START programme.

**Author contributions**: Q.X. and E.A.O. supervised and guided this research. R.S. and Y.H. designed the setup. R.S. fabricated the sample. R.S and E.E designed and performed the experiments. E.E., R.S. and D.B. analysed and interpreted the data with input from M.W., M.P, A.G.T., T.C.H.L, E.A.O., and Q.X.. E.E. developed the theoretical model and performed the theoretical calculations. E.E., R.S., and E.A.O. wrote the manuscript with the input from all authors.

**Competing interests**: The authors declare that they have no competing interests.

**Figures and Tables**

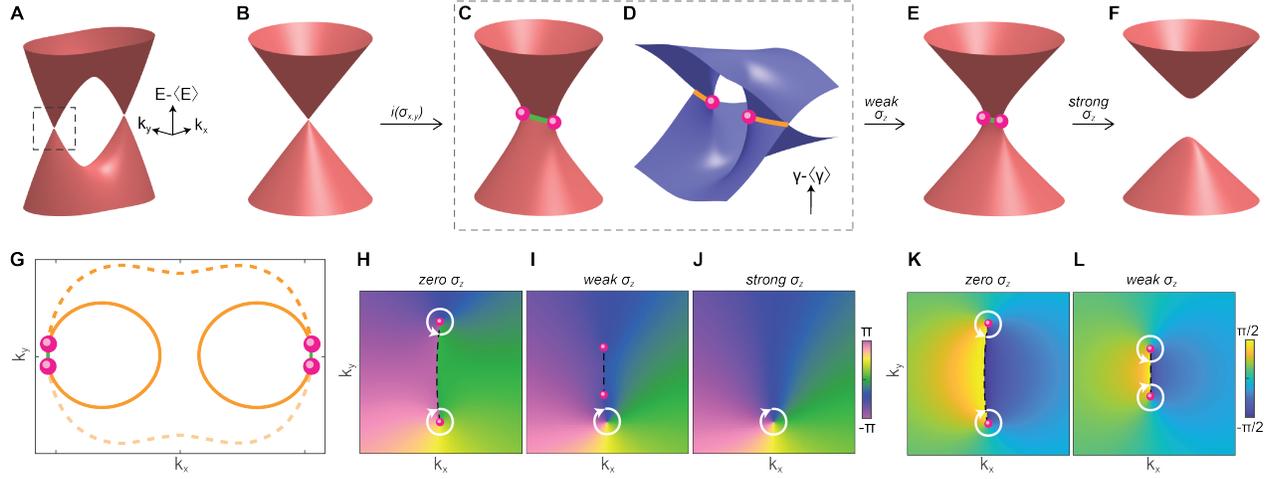

**Fig. 1. Complex spectral structure near pairs of exceptional points in momentum space.** (**A**) Energy (real part of the complex spectrum) of the exciton-polariton modes in a microcavity with linear birefringence, calculated using the model Eq. (1). The mean energy is subtracted for clarity. Energy crossings occur at two opposite regions in the 2D momentum space ($k_x$, $k_y$). (**B**) Enlarged view of the dashed region in **A** in the Hermitian limit, showing a Dirac point. (**C**) Energy of the dashed region in (A) in the non-Hermitian case, with nonzero $i\sigma_{x,y}$ components, showing the Dirac point splitting into a pair of EPs (pink dots) connected by the nodal line - bulk Fermi arc (green), where the energies cross. (**D**) Imaginary part of the complex spectrum corresponding to the linewidth for the dashed region in (A), showing the imaginary Fermi arc (orange), where the linewidths cross, emanating from the exceptional points (pink dots). (**E**) Energy of the system with a weak, real-valued $\sigma_z$-term perturbation. (**F**) Same as E but with a strong perturbation leading to the annihilation of the EPs and opening of the gap. (**G**) Simplified complex energy structure of the two eigenstates, showing the bulk (green) and imaginary (orange) Fermi arcs connecting at the exceptional points and forming two closed contours. A single contour can also form (dashed orange) for the different sign of the parameters in Eq. (3). (**H**, **I**, and **J**) In-plane pseudospin angle in momentum space of the upper eigenstate corresponding to (from left to right) C, E, F, respectively. (**K** and **L**) Spectral phase $Arg(\Delta \tilde{E})$ in momentum space corresponding to C and E, respectively. In (H and I), pink dots correspond to the EP, dashed lines correspond to the bulk Fermi arc, and white arrowed contours correspond to the half-charge (H, K, L) and integer (I, J) windings around the singularities.

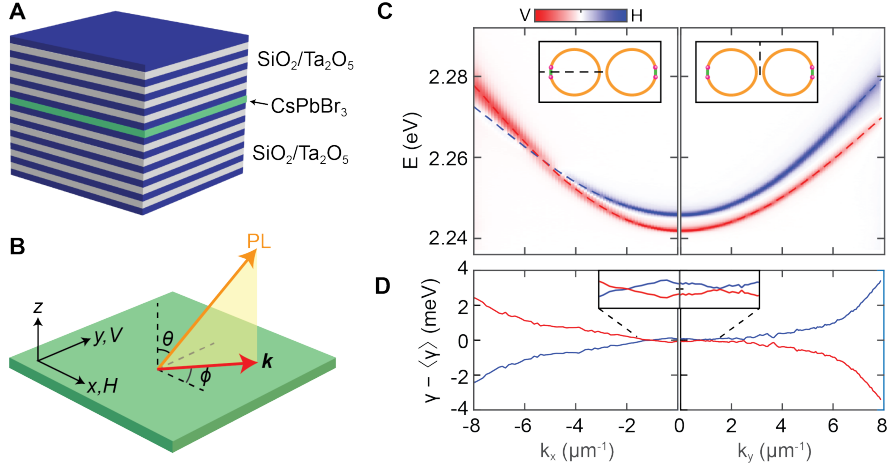

**Fig. 2. Experimental investigation of the complex exciton-polariton eigenenergies. (A)** Schematics of the planar microcavity made of $SiO_2/Ta_2O_5$ DBRs with an embedded $CsPbBr_3$ perovskite crystal. **(B)** Schematics of the laboratory $(x, y, z)$ axis and the polarisation measurement axis $(H, V)$. The exciton-polariton in-plane momentum depends on the angles $(\theta, \phi)$ of the PL emission. **(C)** Linearly polarized PL intensity $(I_V$-$I_H)$ measured along $(k_x, k_y = 0)$ and $(k_x = 0, k_y)$. Dashed lines are the extracted peak energies of the two polarised modes. The dispersion is approximately symmetric for $k \rightarrow -k$. Inset: Schematics of the measurements in momentum space with respect to the Fermi arcs. **(D)** Linewidths of the modes in **C** with the mean subtracted. Inset: enlarged region near $k = 0$.

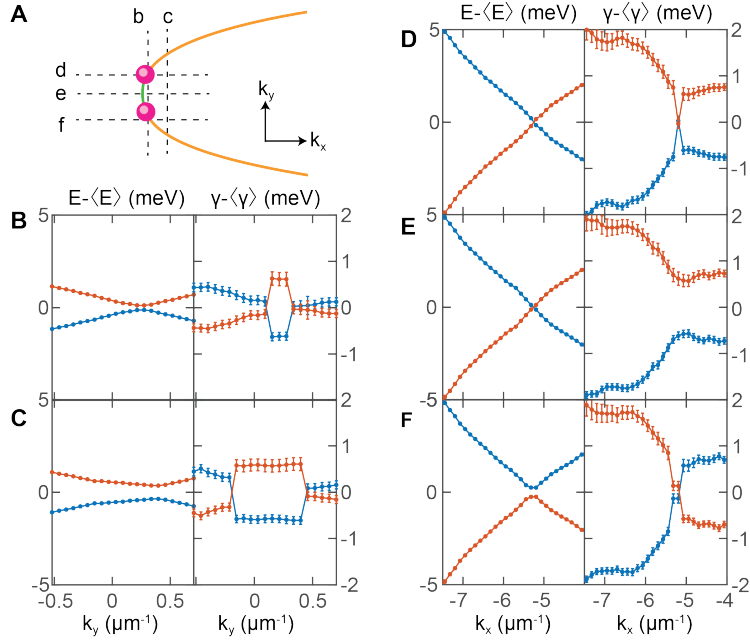

**Fig. 3. Mapping out complex energies near the EP pair.** (**A**) Schematics of the EP pair (pink dots) connected by the bulk (green) and imaginary (orange) Fermi arcs. Dashed lines (b-f) represent the lines (directions) in *k*-space, along which the measurements in (B-F) are performed. (**B-F**) Measured energies and linewidths (mean-subtracted) of the two modes: B, parallel to and very near the bulk Fermi arc; C, parallel to the bulk Fermi arc intersecting the imaginary Fermi arc twice, which corresponds to two linewidth crossings and no crossing in energy; D, perpendicular to the bulk Fermi arc very near the top EP, showing crossing in both energy and linewidth; E, along the centre of the real Fermi arc, showing crossing in energy and anticrossing in linewidth; F, near the EP but outside the real Fermi arc showing no crossing in energy but crossing in linewidth. The complex eigenvalues are sorted so that a smooth crossing (D, E) or anti-crossing (B, C, F) in the real part is ensured. The values for *k* are: (B) $k_x = -5.19$ μm$^{-1}$, (C) $k_x = -5.07$ μm$^{-1}$, (D) $k_y = 0.40$ μm$^{-1}$, (E) $k_y = 0.21$ μm$^{-1}$, (F) $k_y = 0.09$ μm$^{-1}$. Error bars represent the 95% confidence interval fitting results.

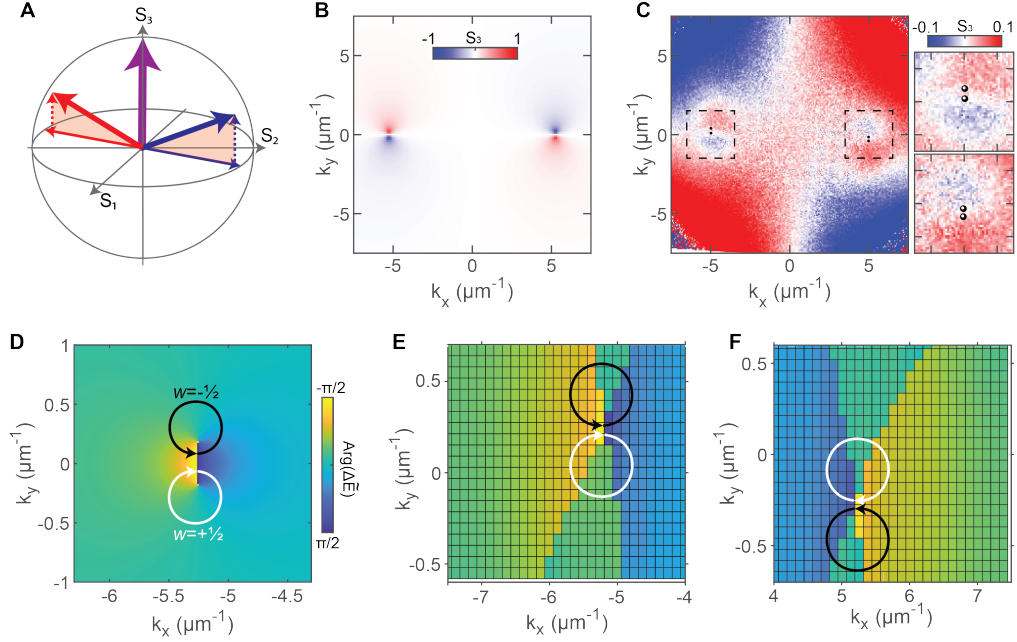

**Fig. 4. Chirality and topology of the exceptional points.** (**A**) Poincare sphere with arrows representing the pseudospin of exciton polaritons away from the EP (thin red and blue), near the EP (thick red and blue), and at the EP (thick purple). Dashed vertical arrows are the effective out-of-plane field arising from the imaginary component of the complex in-plane artificial magnetic field. (**B**) Theoretical texture of circular polarisation ($S_3$) arising from the inclusion of non-Hermiticity into the model of Eq. (1). (**C**) Measured energy-integrated circular polarisation ($S_3$) showing the same spin structure as in B but with a weak $S_3$ background coming from the bare perovskite (see text). Right panels: enlarged images of the marked regions showing the position of EPs (black points). (**D**) Theoretical values of $\arg(\tilde{E}_+ - \tilde{E}_-)$ for one EP pair with the arrows schematically showing the fractional winding number. Parameters are the same as in Fig. 1(C and D). (**E** and **F**) Measured values of $\arg(\tilde{E}_+ - \tilde{E}_-)$ near the two pairs of EPs demonstrating the half-integer spectral winding around each EP.